\input{aipcheck}

\documentclass[
    ,final            
  ]
  {aipproc}

\layoutstyle{6x9}

\begin{document}

\title{
Non-axisymmetric Structure of Accretion Disks in Be/X-ray Binaries}


\author{Kimitake Hayasaki}
{
  address={Department of Applied Physics, Graduate School of Engineering, Hokkaido University, 
           Kitaku N13W8, Sapporo 060-8628, Japan}
  ,altaddress={Centre for Astrophysics and Supercomputing, Swinburne University of Technology, Hawthorn Victoria 3122, Australia},
,email={hayasaki@topology.coe.hokudai.ac.jp}
}
\author{Atsuo T. Okazaki}{
  address={Faculty of Engineering, Hokkai-Gakuen University, Toyohira-ku, Sapporo 062-8605, Japan}
}

\begin{abstract}
The non-axisymmetric structure of accretion disks around the neutron star
in Be/X-ray binaries is studied by analyzing the results 
from three dimensional (3D) Smoothed Particle Hydrodynamics (SPH) simulations.
It is found that ram pressure due to the phase-dependent
mass transfer from the Be-star disk excites
a one-armed, trailing spiral structure
in the accretion disk around the neutron star.
The spiral wave has a transient nature;
it is excited around the periastron,
when the material is transferred from the Be disk,
and is gradually damped afterwards.
It is also found that
the orbital phase-dependence of the mass-accretion rate
is mainly caused by the inward propagation of
the spiral wave excited in the accretion disk.
\end{abstract}

\maketitle


\section{INTRODUCTION}

The majority of high-mass X-ray binaries
have been identified as Be/X-ray binaries.
These systems generally consist of a neutron star and
a Be star with a cool ($\sim 10^{4}K$) equatorial disk, which is
geometrically thin and nearly Keplerian.
Be/X-ray binaries are distributed over a wide range of orbital
periods 
$(10\,{\rm d} \le P_{\rm{orb}} \le 300\,\rm{d})$ and
eccentricities ($e \le 0.9$).

Most Be/X-ray binaries show only transient activity in 
X-ray emission and are termed Be/X-ray transients. Be/X-ray transients
show periodic (Type I) outbursts, which are separated by the orbital
period and have lumiosity $L_{\rm{X}}=10^{36-37}\rm{erg\,s}^{-1}$, and giant (Type II)
outbursts of 
$L_{\rm{X}}\ge10^{37}\rm{erg\,s}^{-1}$ with no
orbital modulation.
These outbursts have features that strongly suggest the presence of
an accretion disk around the neutron star.

\begin{figure}
\centerline{\includegraphics[height=5cm]{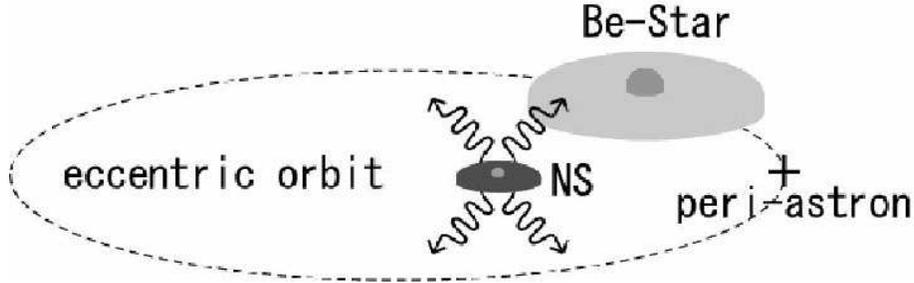}}
\caption
{A new schmatic diagram in a typical Be/X-ray binary.
The system has double circumstellar disks 
(the Be disk around the Be star 
and the accretion disk around the neutron star) which interact mainly via the mass transfer.
}
\label{fig:bex}
\end{figure}

Recently, \citet{kh} studied the
accretion flow around the neutron star in a Be/X-ray binary with a
short period ($P_{\rm orb}=24.3\,{\rm d}$) and a moderate eccentricity
($e=0.34$), using a 3D SPH code 
and the imported data from \citep{oka2002}.
They found that a time-dependent accretion disk is formed around the
neutron star. 
They also discussed the evolution of the
azimuthally-averaged structure of the disk, 
in which a one-armed spiral structure is seen.
It is important to note that Be/X-ray
binaries are systems with double circumstellar disks (the Be
decretion disk and the neutron star accretion disk), which
interact mainly via mass transfer, and that this gives a new
point of view to understand the interactions in Be/X-ray
binaries (see Fig.~\ref{fig:bex}). 

In this paper, we show that 
ram pressure due to the material transferred from the
Be disk around periastron temporarily excites the one-armed spiral wave
in the accretion disk around the neutron star in Be/X-ray binaries.

\section{one-armed spiral wave excited by a ram pressure}

Our simulations were performed by using the same 3D SPH code as in
\citep{kh}, which was based on a version originally developed by Benz
(\citep{benz}; \citep*{benz1}) and later by \citep*{bate}.
In order to study the effect of the ram pressure on the
accretion disk, we compare results from model~1 in \citep{kh} (hereafter,
model~A) with those from a simulation (hereafter, model~B) in which
the mass transfer from the Be disk is artificially stopped for one
orbital period. Except for this difference, the two simulations have the
same model parameters: The orbital period $P_{\rm orb}$ is
24.3\,d, the eccentricity $e$ is 0.34, and the Be disk is coplanar
with the orbital plane. The inner radius of the simulation region
$r_{\rm{in}}$ is $3.0\times10^{-3}a$, where $a$ is the semi-major axis
of the binary. The polytropic equation of state with the exponent
$\Gamma=1.2$ is adopted. The Shakura-Sunyaev viscosity parameter
$\alpha_{\rm{SS}}=0.1$ throughout the disk.

\begin{figure}
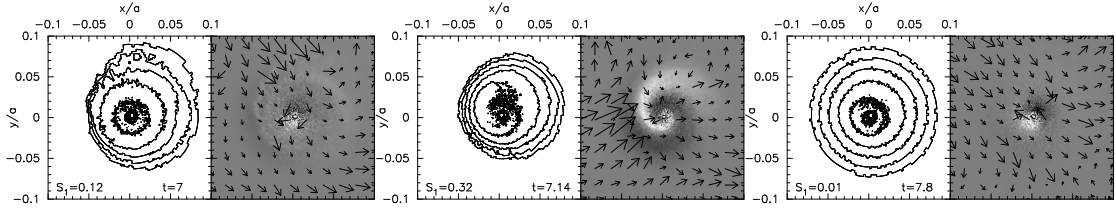

\resizebox{\hsize}{!}{
  \includegraphics*{hayasaki2}\\
  \includegraphics*{hayasaki3}\\
  \includegraphics*{hayasaki4}}
  \caption{Snapshots of the accretion disk for model~A.
   The left panels show the surface density in a range of
   three orders of magnitude in the logarithmic scale,
   while the right panels show the non-axisymmetric components of
   the surface density (gray-scale plot)
   and the velocity field (arrows) in the linear scale.
   In the right panels, the region in gray (white) denotes the region
   with positive (negative) density enhancement.
   The periastron is in the $x$-direction and
   the disk rotates counterclockwise.
   Annotated in each left panel are the time
   in units of $P_{\rm orb}$ and the mode strength $S_{1}$.}
 \label{fig:contours}
\end{figure}

Fig.~\ref{fig:contours} gives a sequence of snapshots of the accretion
disk around the neutron star for $7\le{t}\le8$ in model~A,
where the unit of time is the orbital period $P_{\rm orb}$.
The left panels show the contour maps of the surface density,
whereas the non-axisymmetric components of the surface density
and the velocity field are shown in the right panels.
Annotated in each left panel are the time in units of $P_{\rm orb}$
and the mode strength $S_{1}$, a measure of the amplitude of
the one-armed spiral wave, which is defined by using the azimuthal Fourier decomposition
of the surface density distribution \citep[see][Sec~2.2 for detail]{kh1}.
It is noted from the figure
that the one-armed, trailing spiral is excited at periastron
and is gradually damped towards the next periastron.
The disk changes its topology from circular to eccentric 
with the development of the spiral wave, and then
from eccentric to circular with the decay of the wave 
during one orbital period.

\begin{figure}
\centerline{\includegraphics[height=2.65cm]{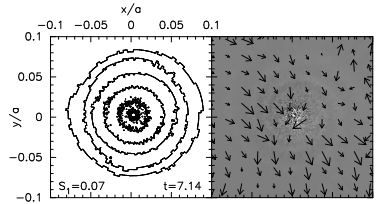}}
	\caption{Same as Fig. 2, but for model~B.}
 \label{fig:contours2}
\end{figure}

For comparison, we present the results for model~B, in which
the mass transfer is artificially turned off for $7 \le t \le 8$.
Fig.~\ref{fig:contours2} shows
the surface density (the left panel)
and the non-axisymmetric components of the surface density
and the velocity field (the
right panel) at the time corresponding to the middle panel of
Fig.~\ref{fig:contours}. 
It should be noted that
disk deformation due to the one-armed mode is not seen in model~B.
The disk is more circular and has a larger radius in model~B than in
model~A.
This strongly suggests that
the excitation of the one-armed spiral structure in the accretion disk
is induced by the ram pressure from the material transferred from the
Be disk at periastron.

\subsection{Phase dependence of the mass-accretion rate}
\label{sec:accrate}

After the accretion disk has developed ($t \ge 5$), 
the mass-accretion rate has double peaks per orbit,
a relatively-narrow, low peak at periastron and a broad, high peak afterwards
[see Fig.~15(a) of \citep{kh}].
While the first low peak at periastron could be artificial, being related to
the presence of the inner simulation boundary,
the origin of the second high peak was not clear.
Below we show that the one-armed spiral wave is responsible for the second peak
in the mass-accretion rate.

\begin{figure}
\centerline{\includegraphics[height=10cm]{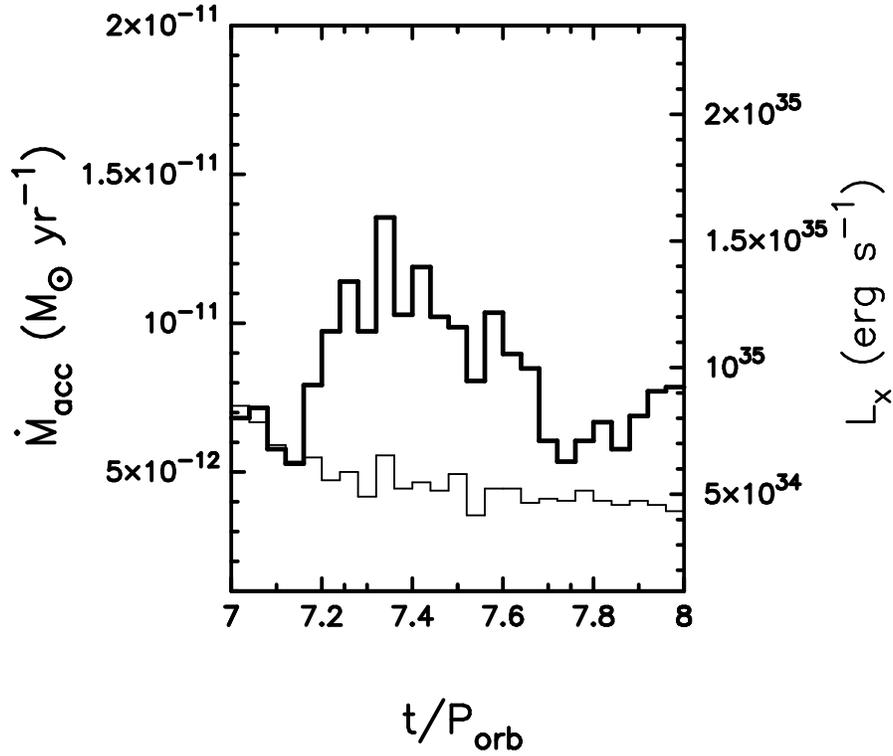}}
	\caption{Time dependence of the mass-accretion rate for $7\le{t}\le8$.
	The thick and thin lines are for model~A and model~B, respectively.
	The right axis shows the X-ray luminocity corresponding
	to the mass-accretion rate.}
\label{fig:accrate}
\end{figure}

Fig.~\ref{fig:accrate} shows
the time dependence of the mass-accretion rate for $7\le{t}\le8$.
The thick line denotes the mass-accretion rate in
model~A, in which the mass transfer from the Be disk 
is taken into account, whereas the thin line denotes that in model~B. 
The difference between the accretion rate profiles for these two models is striking.
The accretion rate in model~B monotonically decreases over one orbital period,
whereas that of model~A shows a broad peak centred at $t\sim7.32-7.35$,
which corresponds to the second peak found in \citep{kh}.
The phase lag of this peak behind the peak of the mass-transfer rate, which occurs at periastron, 
results from the inward propagation of the wave 
from the disk outer radius to the inner simulation boundary.

\begin{theacknowledgments}
This work has been supported by Grant-in-Aid for the 21st Century 
COE Scientific Research Programme on "Topological Science and Technology"
from the Ministry of 
Education, Culture, Sport, Science and Technology of Japan (MECSST) and
in part by Nukazawa Science Fundation.
\end{theacknowledgments}


\end{document}